\documentclass[hidelinks, conference]{IEEEtran}
\makeatletter
\def\ps@headings{%
\def\@oddhead{\mbox{}\scriptsize\rightmark \hfil \thepage}%
\def\@evenhead{\scriptsize\thepage \hfil \leftmark\mbox{}}%
\def\@oddfoot{}%
\def\@evenfoot{}}
\makeatother
\newcommand*{\longDefiningEquals}{\stackrel{\text{def}}{=\joinrel=}}

\usepackage{color}
\usepackage{url}
\usepackage[latin9]{inputenc}
\usepackage{amsthm}
\usepackage{amsmath}
\usepackage{amsthm}
\usepackage{amssymb}
\usepackage{graphicx}
\usepackage{epstopdf}
\usepackage{wrapfig}
\usepackage{algorithmic}
\usepackage{algorithm}
\usepackage{mathrsfs}
\usepackage{float}
\usepackage{mathtools}
\usepackage{tabulary}
\usepackage{booktabs}
\usepackage{caption}
\usepackage{subfig}
\usepackage{xprintlen}
\usepackage{multirow}
\usepackage{textcomp}

\graphicspath{{./}{./files/figures_eps/}}

\PassOptionsToPackage{bookmarks={false}}{hyperref}

\IEEEoverridecommandlockouts
\usepackage{graphicx}
\usepackage{svg}

\def\BibTeX{{\rm B\kern-.05em{\sc i\kern-.025em b}\kern-.08em
   T\kern-.1667em\lower.7ex\hbox{E}\kern-.125emX}}

\newcommand{\comment}[1]{ }

\newcommand\subparagraph{%
  \@startsection{subparagraph}{0}
  {\parindent}
  {0ex \@plus 0ex \@minus 0ex}
  {-1em}
  {\normalfont\normalsize\bfseries}}
\makeatother

\usepackage{titlesec}
\titlespacing*{\section}
{0pt}{1.2ex}{0ex} 
\titlespacing*{\subsection}
{0pt}{1ex}{0ex}  
\titlespacing*{\subsubsection}
{0pt}{0ex}{0ex} 

\begin{document}











\title{Domain-Adaptive Device Fingerprints for Network Access Authentication Through Multifractal Dimension Representation}


\author{Benjamin Johnson and Bechir Hamdaoui~\\
School of Electrical Engineering and Computer Science~\\
 Oregon State University, Corvallis, OR, USA ~\\ 
 \{johnsbe3,hamdaoui\}@oregonstate.edu
\thanks{This work is supported in part by NSF/Intel Award No. 2003273.}
}

\maketitle
\thispagestyle{plain}
\pagestyle{plain}

\begin{abstract}
RF data-driven device fingerprinting through the use of deep learning has recently surfaced as a potential solution for automated network access authentication. Traditional approaches are commonly susceptible to the domain adaptation problem where a model trained on data from one domain performs badly when tested on data from a different domain. Some examples of a domain change include varying the device location or environment and varying the time or day of data collection. In this work, we propose using multifractal analysis and the variance fractal dimension trajectory (VFDT) as a data representation input to the deep neural network to extract device fingerprints that are domain generalizable. We analyze the effectiveness of the proposed VFDT representation in detecting device-specific signatures from hardware-impaired IQ signals, and evaluate its robustness in real-world settings, using an experimental testbed of 30 WiFi-enabled Pycom devices under different locations and at different scales. Our results show that the VFDT representation improves the scalability, robustness and generalizability of the deep learning models significantly compared to when using raw IQ data. 
\end{abstract}
 

\section{Introduction}
\label{intro}
Wireless device identification and authentication through the use of RF (radio frequency) fingerprinting has recently been considered as a new potential method for enabling authenticated network access without requiring implicit trust from any network entity~\cite{gul2023secure,hamdaoui2023deep,8715341,elmaghbub2023eps,liu2023robust}. 
In essence, RF-based device fingerprinting consists of extracting hardware-impaired, device-specific signatures using only the raw RF signals transmitted by the wireless devices. These hardware impairments typically come from the inherent `random' variability in the manufacturing process of the devices, and therefore, they are difficult to reproduce or replicate, making such methods robust against impersonation attacks.
The commonly used methods of extracting the fingerprints and classifying the devices mostly rely on deep learning models, which need to be trained first on labeled RF data, and then used to identify and authenticate devices~\cite{jagannath2022comprehensive,basha2023channel,one_DeepRadioID2019,puppo2023hinova}.

While deep learning models have already been shown to offer promising results, they come with some inherent limitations and problems~\cite{zhang2023radio,hamdaoui2022deep,mackey2022cross,elmaghbub2023adl}. 
Due to the design nature of these learning models and how deep neural networks are trained, the exact features being used to identify and distinguish devices are unknown. This essentially means the deep learning model is a black box and as such could be using something other than the inherent hardware variations to classify devices. For instance, these models could be focusing on the channel conditions instead of the hardware impairments which would lead to incorrect device identification and classification when the model is used under different channel conditions. This limitation of the models not being able to generalize to other conditions and settings is often referred to, in the RF fingerprinting community, as the {\em domain adaptation} or {\em model generalization} problems. 

\begin{figure}[t]
\centering
\includegraphics[width=\columnwidth]{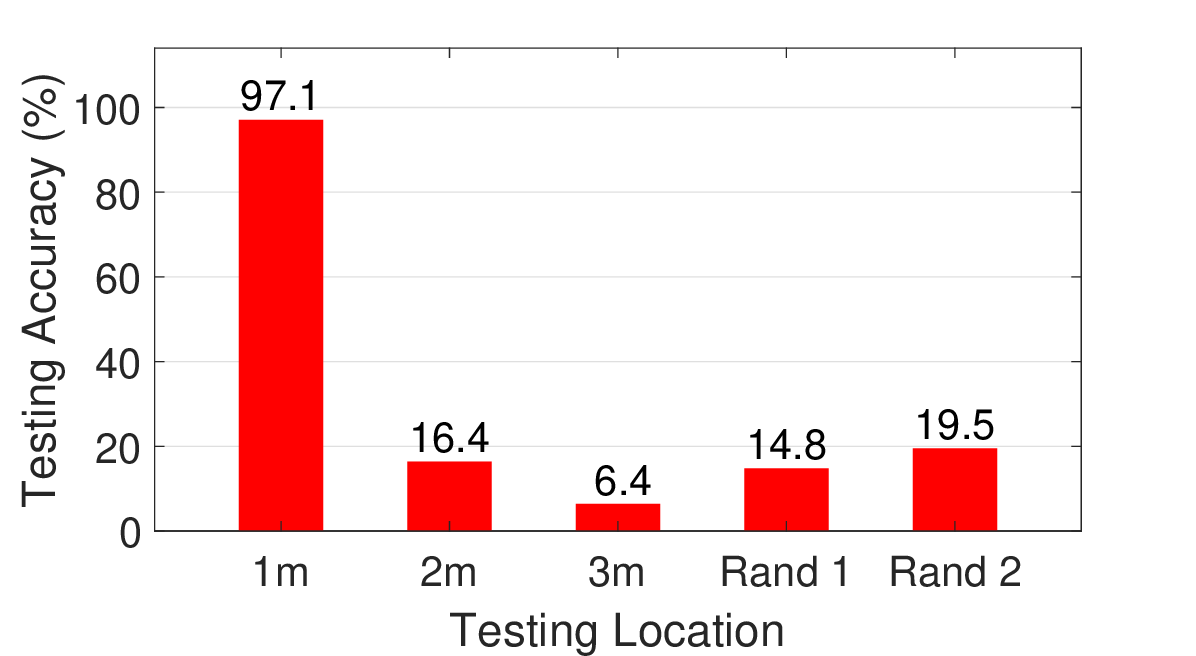}
\caption{CNN model classifier trained on data collected from 30 devices placed 1m away from the receiver, but tested on data collected when the devices are placed at different distances from the receiver: 1m, 2m, 3m, and two random locations. Experiments are taken indoor, in a lab environment.}
\label{fig:motivation}
\end{figure}

It has been shown in several studies (e.g.~\cite{gaskin2022tweak,four_scalableLoRa2022,two_deepLoRa2021}) that deep learning models are able to classify devices with very high accuracy when both the testing data and training data are taken under the same domain (e.g., channel, time, receiver, location). However, when these models are tested on data taken under a different domain (e.g., training and testing data are collected under different channel conditions), their accuracy is greatly reduced. To demonstrate this domain-adaption challenge, we collected RF datasets using our experimental testbed, consisting of 30 Pycom devices (more details on the testbed are provided in Sec.~\ref{sec:testbed}), and used the datasets to train and test a classical CNN (Convolution Neural Networks) classifier, while considering varying distances between the transmitters and the receiver during training and testing, thereby varying the wireless channel.
The results of this experiment displayed in Fig.~\ref{fig:motivation} show the testing accuracy of the deep learning model, trained on data collected when the devices are placed 1m away from the receiver but tested on data collected when the devices are placed at different locations, 1m, 2m, 3m and at random.
These results show a significant drop in accuracy when the model is tested on data taken from a different domain, in this case the physical location. Note that when the training and testing are both done when the devices are 1m away from the receiver, the testing accuracy is above 97\%. However, due to the limitation mentioned above; i.e., the inability of the learning model to adapt to domain changes, when the model is trained on data collected 1m away but tested on data collected 2m away, the testing accuracy drops from 97\% to only about 16\%. Our experimental results clearly demonstrate the severity of the domain-adaptation problems in RF data-driven device fingerprinting, and thus the need for coming up with novel approaches that overcome such problems. 

There have been some attempts to address these domain-adaptation problems \cite{four_scalableLoRa2022,gaskin2022tweak,9771750,10012578,3529520}. 
Some attempts proposed data preprocessing approaches to help extract domain-independent fingerprints that can improve classification accuracy when fed to deep learning models~\cite{four_scalableLoRa2022}. Others attempted to directly calibrate the model itself through few-shot samples to overcome portability problems across domains \cite{gaskin2022tweak}. Another potential approach is through the use of multifractal analysis to extract the innate hardware impairments and use them for device classification~\cite{10012578}. However, these attempts do not examine how their identification results are affected by domain adaptation or how scalable their proposed approaches are to the number of tested devices. Unlike those efforts, our work proposes to leverage multifractal signal analysis to extract device-specific, domain-independent fingerprints that are both scalable and robust to domain changes.



This paper leverages multifractal signal analysis to extract hardware-impaired device features from received RF signals and present them as an input to the deep learning classifiers. Specifically, the proposed method involves capturing the raw IQ samples at the receiver and separately calculating the variance fractal dimension trajectory (VFDT) of both the in-phase (I) and quadrature (Q) components. The resulting VFDT output signals are then fed to a CNN-based deep learning model to perform the device identification task.
We begin by analyzing the effectiveness of the proposed VFDT representation in capturing device-specific signatures, caused by different hardware-impaired distortions in the RF signals, through real WiFi datasets as well as simulated datasets. We then assess the ability of the proposed method in identifying and classifying wireless WiFi devices, as well as its robustness to domain changes, and we do so using an experimental testbed of 30 WiFi-enabled Pycom devices. 
Our results show that the proposed VFDT representation enhances the scalability, robustness and generalizability of the deep learning models significantly compared to raw IQ data representation. 


The remainder of this work is as follows: Sec.~\ref{sec:background} provides some background on multifractal analysis. Sec.~\ref{sec:impairments} examines the effect of hardware impairments on the VFDT behavior. Sec.~\ref{sec:proposed} presents our device classification method. Sec.~\ref{sec:testbed} describes the testbed and data collection used to evaluate the proposed method, and Sec.~\ref{sec:results} evaluates and analyzes the effectiveness of our method. Finally, we conclude the paper in Sec.~\ref{sec:conc}.






%

\section{Variance Fractal Dimension Trajectory}
\label{sec:background}
Multifractal analysis uses the fractal dimension to characterize how a signal varies or meanders over different scales of measurements. It has been used in various real-world applications ranging from noise estimation~\cite{117957} to fish trajectory analysis~\cite{1225951} to estimating the length of coastlines \cite{1532616}. In general, the fractal dimension can be seen as a representation of the degree of irregularity, complexity, or meandering  of an object or signal~\cite{multifractal}. There are many different classes of fractal dimensions such as morphological fractal dimensions, entropy-based fractal dimensions, and transform-based fractal dimensions~\cite{1532616}. In this paper, we focus on a specific type of transform-based fractal dimensions, known as the variance fractal dimension. This section shows how this type of signal analysis can be used to extract the hardware-impaired fingerprints of wireless devices from received RF signals.

\subsection{The VFDT Data Representation}
Our data representation proposed for enabling efficient fingerprint extraction involves estimating the variance fractal dimension of the RF signals, which is done by analyzing the statistical variance of the signal amplitude over different scales and ranges. 
The variance fractal dimension, \(D\), can be expressed as $E + 1 - H$ where $E$ is the Euclidean dimension and $H$ is the Hurst exponent~\cite{1225951}.
%
%
For the case of RF data signals, after the in-phase and quadrature components are separated, the Euclidean dimension $E=1$, and thus, the variance fractal dimension $D=2 - H$ (see~\cite{1225951} for details).

\comment{
Our proposed data representation for enabling efficient fingerprint extraction involves estimating the variance fractal dimension of the RF signals using multifractal analysis. This variance dimension is based on analyzing the statistical variance of the signal amplitude over different scales and ranges. 
In order to estimate this variance dimension, it is first assumed that the signal adheres to the power law relationship, 
\begin{equation}\nonumber
   \operatorname{var}[x(t_2)-x(t_1)] \sim |t_2 - t_1|^{2H},
\end{equation}
over any time interval $t_2 - t_1$, where \(x(t)\) is the signal at time $t$ and \(H\) is the Hurst exponent \cite{1532616}. 
The variance fractal dimension \(D\) is then equal to $E + 1 - H$,
where \(E\) is the Euclidean dimension \cite{1225951}. 
For the case of RF signals, after the in-phase and quadrature components are separated, we have $E=1$ (one-dimensional) or $D = 2 - H$, where $D$ again represents the fractal dimension of the overall signal.
}%

Our observation here is that for the time-varying RF signals, it is a more useful representation to capture the time-variability of the fractal dimension, and as such, we propose to use the Variance Fractal Dimension Trajectory (VFDT), which is a rolling trajectory of the fractal dimension, to represent the IQ data that is to be fed to the deep learning classifiers. Later in the paper, we will show the effectiveness of this representation on addressing the domain-adaption problem of device fingerprinting. But now in this section, we focus on analyzing and understanding how effective VFDT of the IQ signals, as a data representation, is in distinguishing and separating between devices. Before doing so, let's first describe and explain how we go about calculating VFDT of a time-varying signal, like the I and Q signals.
 
To calculate VFDT of a discrete-time signal, the fractal dimension, \(D\), is calculated for a windowed segment of the sampled signal. This windowed segment is then shifted or offset by a fixed, predetermined amount of samples, which is then used for calculating the fractal dimension. The process is repeated until the end of the signal samples is reached. Due to the signal being discrete, the length of the windowed segment, $\Delta w$, is used as the time interval, and VFDT of a given windowed segment \(i\) of size $\Delta w$ can be estimated as \cite{1225951}
\begin{equation} 
    \operatorname{VFDT}(i) \longDefiningEquals D(i)= 2 - {\log[\operatorname{var}(\Delta x)]}/{(2\log(\Delta w))}
    \label{eg:vfdt}
\end{equation}
with $\Delta x$ being the signal difference evaluated at the beginning and end of the interval $\Delta w$. The length of the windowed segment along with the size of the window offset used can both affect the resulting VFDT. The best window length and window offset will be unique and specific for any given type of signal. For the RF signals analyzed in this paper, the main factors that resulted in needing to vary the window size and offset seemed to be the sample rate of the signal along with the stationarity of the signal. Generally, the best window length and window offset are subjective and commonly determined through experimentation. We found it useful to have the window offset size be smaller than the window length, as this causes an overlap in adjacent windows and typically yields better results. Also note that the smaller the window length and window offset are, the larger the resulting VFDT will be in terms of analyzed segments.

\subsection{VFDT Separability Across Different WiFi Devices}
Our objective now is to assess VFDT's ability in separating between devices. For this, we first use a USRP receiver to collect IQ signals that are sent by multiple different WiFi-enabled Pycom devices. Then, we apply and visualize the VFDT on each of the received IQ signals to see how separable the calculated VFDTs are across different devices. 

\begin{figure}[t!]
\centering
\includegraphics[width=\columnwidth]{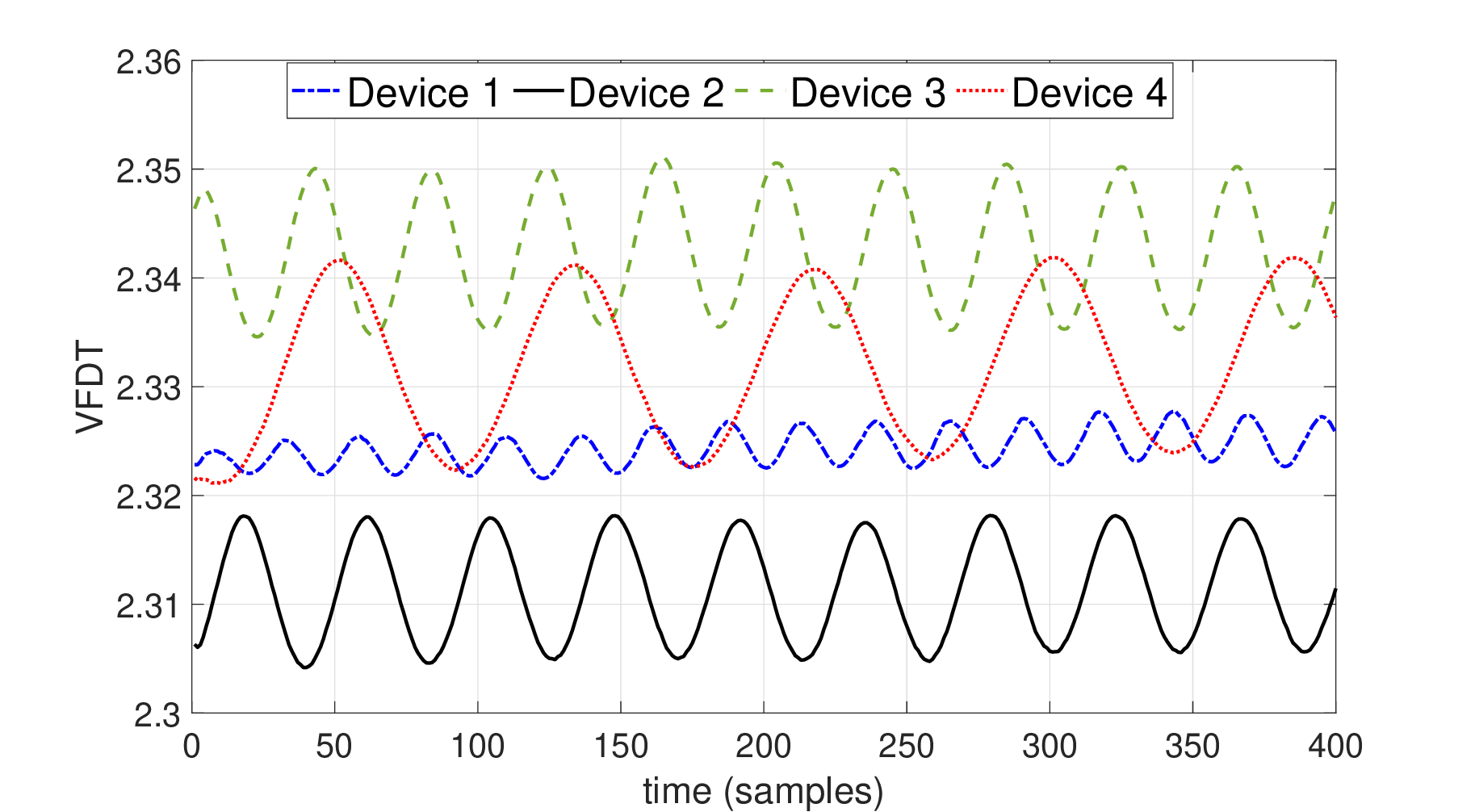}
\caption{VFDT of the I components of the IQ signals collected from 4 different  WiFi-enabled LoPy/Pycom devices.}
\label{fig:wifi}
\end{figure}

We collected IQ data from 10 different WiFi-enabled LoPy devices, transmitting at the 2.4GHz band. The transmissions were sampled at 45MS/s. Each LoPy device was placed 1m away from the receiver in an indoor environment and transmitted data was collected for 2 minutes. More testbed details can be found in Sec.~\ref{sec:testbed}. 
Plots of the calculated VFDT values for the I component of 4 different WiFi-enabled devices are shown in Fig. \ref{fig:wifi}. Observe that the VFDT values of all 4 devices are consistent across the entire frame and have values that are well distinct across the different devices. One interesting result of note is how each device's VFDT seems to oscillate and at a certain frequency. The oscillating behavior is likely due to the presence of carrier frequency offset between the transmitter and the USRP receiver, which differs from one device to another, resulting in different frequency of the sinusoidal shape~\cite{elmaghbub2023impact}.
%
These VFDT results further suggest that the combination of hardware impairments for each device creates a unique fingerprint that is well captured in the VFDT representation.

\section{Hardware Impairments and Their Impact on the VFDT of the Received IQ Signals}
\label{sec:impairments}
\subsection{Transceiver Hardware Impairments}
A wireless device's RF transceiver consists of various hardware components, including digital-to-analog converters, local oscillators, mixers, and power amplifiers. Each component comes with undesired hardware impairments that result from the manufacturing process. These impairments manifest themselves in various distortions, including IQ imbalance, DC offset, phase noise, carrier frequency offset, and power amplifier nonlinearity~\cite{razavi2012rf,elmaghbub2021Lora}, and vary across devices, resulting in each device having a unique set of distinctive fingerprints that can be used to uniquely identify it. 
While this fingerprint or signature is embedded in the transmitted RF signal, as explained in the introduction section, when the raw RF data is used to train a deep learning model, the embedded fingerprint is typically not captured well enough to correctly classify a device across different domains. Thus, the goal of using multifractal analysis and the proposed VFDT is to allow deep learning models to better extract device-specific fingerprints so as to maintain performance consistency across changing domains.



In the previous section, we showed, through experimentation, the ability of the proposed VFDT representation in capturing device-specific features and in separating among different devices. However, in these real, device-generated signals, the aggregation of all the impairments is what is captured via VFDT. As such, only the overall effect of the impairments as a whole was analyzed with VFDT for these real devices. Given that each type of hardware impairment makes its own contribution to the larger overall distortion of the  signal, in this section, we turn our attention to examining the effect of each impairment on the VFDT behavior (and hence on the device fingerprint). Since it is not possible to vary and adjust the value of a hardware impairment on a real device, here we rely on simulation to do so, allowing us to study and observe the effect of each impairment on the VFDT behavior. 

\subsection{VFDT Separability Under Different Impairments}
We consider studying three key hardware impairments:
power amplifier (PA) nonlinearity, IQ imbalance, and phase noise. 
All of the simulations in this section are performed using MATLAB's predefined impairment models. For each impairment, a random 16,000 bit payload of data is generated. 
Then the 4-QAM modulation is used to digitally modulate the signal, yielding the two in-phase (I) and the quadrature (Q) signal components. The resulting IQ signal is then passed through the specific model for a given impairment. Finally, the received IQ signal output is sampled and analyzed using VFDT. 

\subsubsection{Power Amplifier (PA) Nonlinearity Distortion}
PAs amplify the power of the modulated RF signal prior to its transmission on the antenna. 
For efficiency reasons, PAs often operate in their nonlinear region, causing a nonlinearity distortion that is mainly a result of the amplitude and phase output responses due to changes in the input signal \cite{1231065}. 
When the PA is linear, only the first order coefficient contributes to the output. However, when the PA is nonlinear, the remaining coefficients also contribute to the signal distortion. Typically, the even order coefficients cancel out leaving the third order coefficient as the main cause of distortion. We use MATLAB's implementation, called the cubic polynomial memoryless nonlinear model, to simulate varying degrees of PA nonlinear distortions and analyze their resulting VFDT values.

\begin{figure}
\centering
\includegraphics[width=\columnwidth]{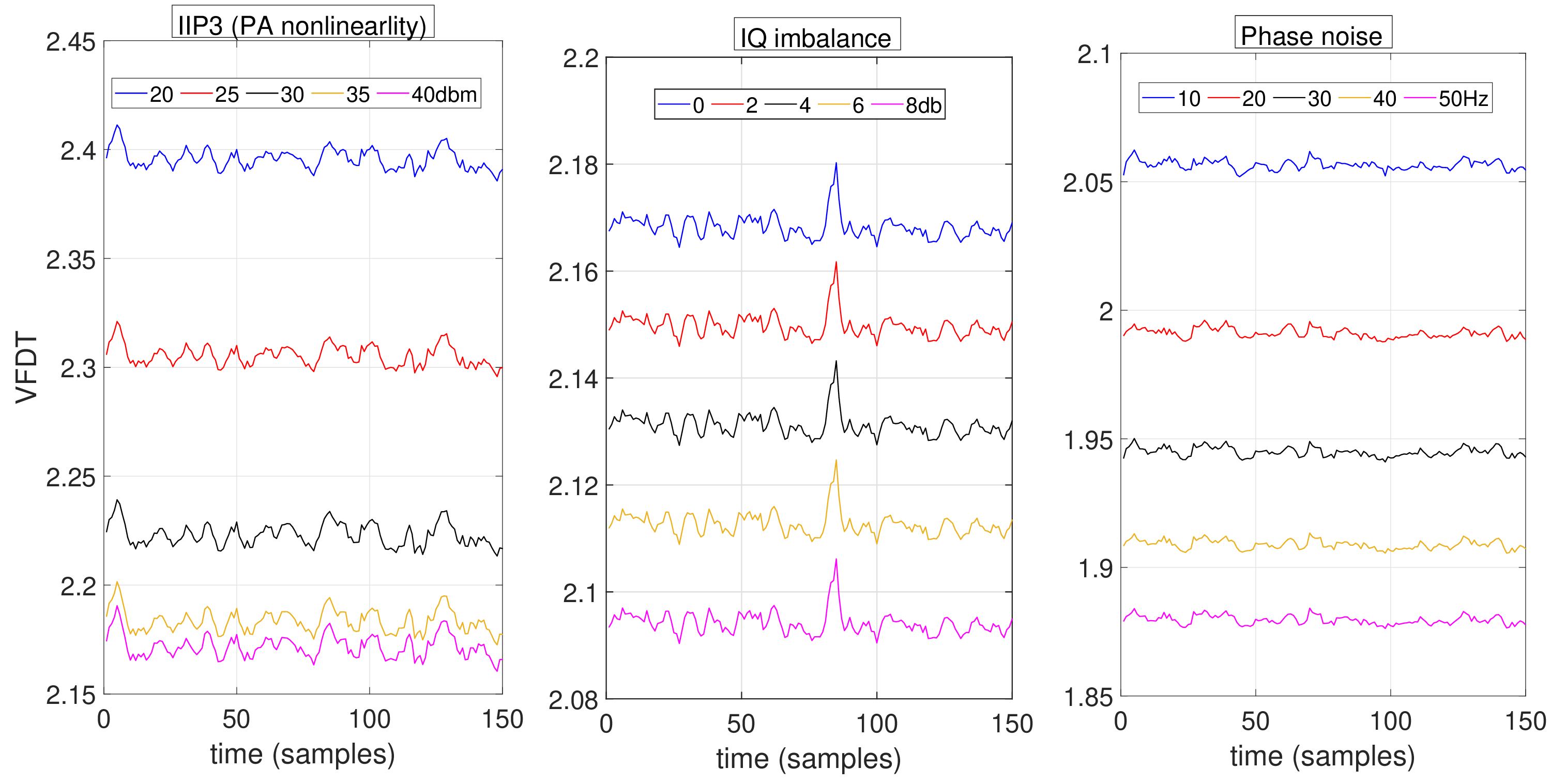}
\caption{VFDT of the I component across different impairments: PA nonlinearlity distortion, IQ mismatch and Phase noise.}
\label{fig:imp}
\end{figure}

In order to vary the amount of distortion caused by PA nonlinearity, we change the third order intercept coefficient (IIP3) parameter, which is a measure of the third order distortion described above. 
We considered 5 different IIP3 levels: 20dBm, 25dBm, 30dBm, 35dBm, and 40dBm. 
VFDT is calculated on the I component of the resulting signal and is shown in Fig.~\ref{fig:imp} (left). The figure clearly shows that different IIP3 values yield large separation in the VFDT representation, demonstrating that using VFDT to capture PA nonliearity distortions provides very distinct device fingerprints that can be used to uniquely identify and distinguish devices.
The same trend has also been observed for the Q component of the IQ signals but not included in the paper to limit redundancy.
%
%

\subsubsection{IQ Imbalance}
In typical transmitters, the I and Q signals are both upconverted to the carrier frequency at the same time with two mixers and a 90\textdegree{} phase shifter for the Q component. If the mixers are ideal and matched, then there is no imbalance between the I and Q components, resulting in a clean complex output signal. However, for real mixers with some mismatch, there will be a
deviation in the amplitude and the phase between the I and Q components \cite{elmaghbub2021Lora},
often referred to as IQ imbalance or mismatch. A mismatch in the phase and/or the amplitude results in distortions of the output signal. 

For simplicity, in our experiments, only the amplitude imbalance parameter of the model is varied to mimic the distortion resulting from the IQ imbalance. For this, we also considered 5 different amounts of IQ imbalance obtained by setting the amplitude imbalance to 0dB, 2dB, 4dB, 6dB, and 8dB. We show in Fig.~\ref{fig:imp} (center) the VFDT of the I component under the 5 studied IQ imbalances. The figure clearly shows direct separation of the VFDT of the I components across the different IQ imbalances. 
This suggests that the VFDT is distinct across devices due to variation among their IQ imbalances, again confirming that the VFDT representation can serve as a good, distinctive device fingerprint. The same trend has also been observed for the Q component but not included in the paper so as to limit redundancy.


\subsubsection{Phase Noise}
Local oscillators (LO) produce the carrier frequency that is used to upconvert the I and Q signals. Ideally, an LO produces a pure sinusoidal wave of a specific frequency. However, in real LOs there are random phase fluctuations that cause the frequency to drift, resulting in an expansion of the signal's spectrum in both sides of the carrier frequency \cite{847872,elmaghbub2021Lora}. This phase deviation or fluctuation is often known as the phase noise.

We used the built-in MATLAB model to vary the phase noise, which implements filtered Gaussian noise to model the slight frequency variations. The frequency offset parameter, which determines the maximum frequency offset possible, is changed in order to vary the amount of phase noise. We also used 5 different phase noise levels with the maximum frequency offset being set to 10Hz, 20Hz, 30Hz, 40Hz, and 50Hz. VFDT is again taken on the resulting I signals and plotted in Fig.~\ref{fig:imp} (right). 
The figure shows that different phase noises can be well separated when captured via VFDT, implying that the proposed VFDT can be used as an effective representation for providing strong separation among the device fingerprints.
 %


\section{The Proposed Classification Method}
\label{sec:proposed}
The proposed technique consists of processing and extracting the hardware impairments from the IQ signals using the VFDT as the input representation to the deep learning model of the RF fingerprinting methods. 
A device's RF transmissions are captured by the receiver and split into its in-phase (I) and quadrature (Q) components, and the VFDT of each component is computed by calculating the variance fractal dimension for every window or segment of the signal using Eq.~\eqref{eg:vfdt}. For each window, the fractal dimension is estimated by first finding the statistical variance of the signal amplitude across all samples in the given window. Next, the log of the variance is taken and divided by the log of the total number of samples in the given window. Finally, the resulting value is scaled and offset to match the definition of the fractal dimension. The computation starts at the beginning of the signal and the window is then shifted by a set predefined amount until the end of the signal is reached. The computed VFDT is a sequence of values representing the variance fractal dimension along different points of the input signal. These two VFDT sequences for the I and Q components are then fed to a CNN-based deep learning model, which outputs the final device classification. 

The CNN architecture used in this work is based off of the CNN described in \cite{elmaghbub2021Lora}, and is implemented using the "PyTorch" library which is based on the Python programming language. The architecture consists first of 6 convolutional blocks which are all made up of 4 different layers. The first layer is a 2D convolutional layer, followed by a batch normalization, leaky ReLU, and max-pooling layer, in that order. After the convolutional blocks, there is a sequence of 3 fully connected blocks used to prevent overfitting and help format the network to have the appropriate final outputs. Each fully connected block contains 3 actual layers. The first is a true fully connected linear layer, followed by a dropout and leaky ReLU layers. Finally, there is a single fully connected linear layer to obtain the proper number of output nodes to classify the corresponding number of devices. The CNN takes in a pair of vectors as input. The input vectors are the resulting VFDTs of the I and Q samples, respectively. Separate VFDTs are taken for the I samples and the Q samples and the resulting vectors are fed as inputs to the CNN in the form of 2x1024.

\section{Testbed and Dataset Collection}
\label{sec:testbed}
\begin{figure}
\centering
\subfloat[B210 receiver]{
    \includegraphics[width=.295\columnwidth]{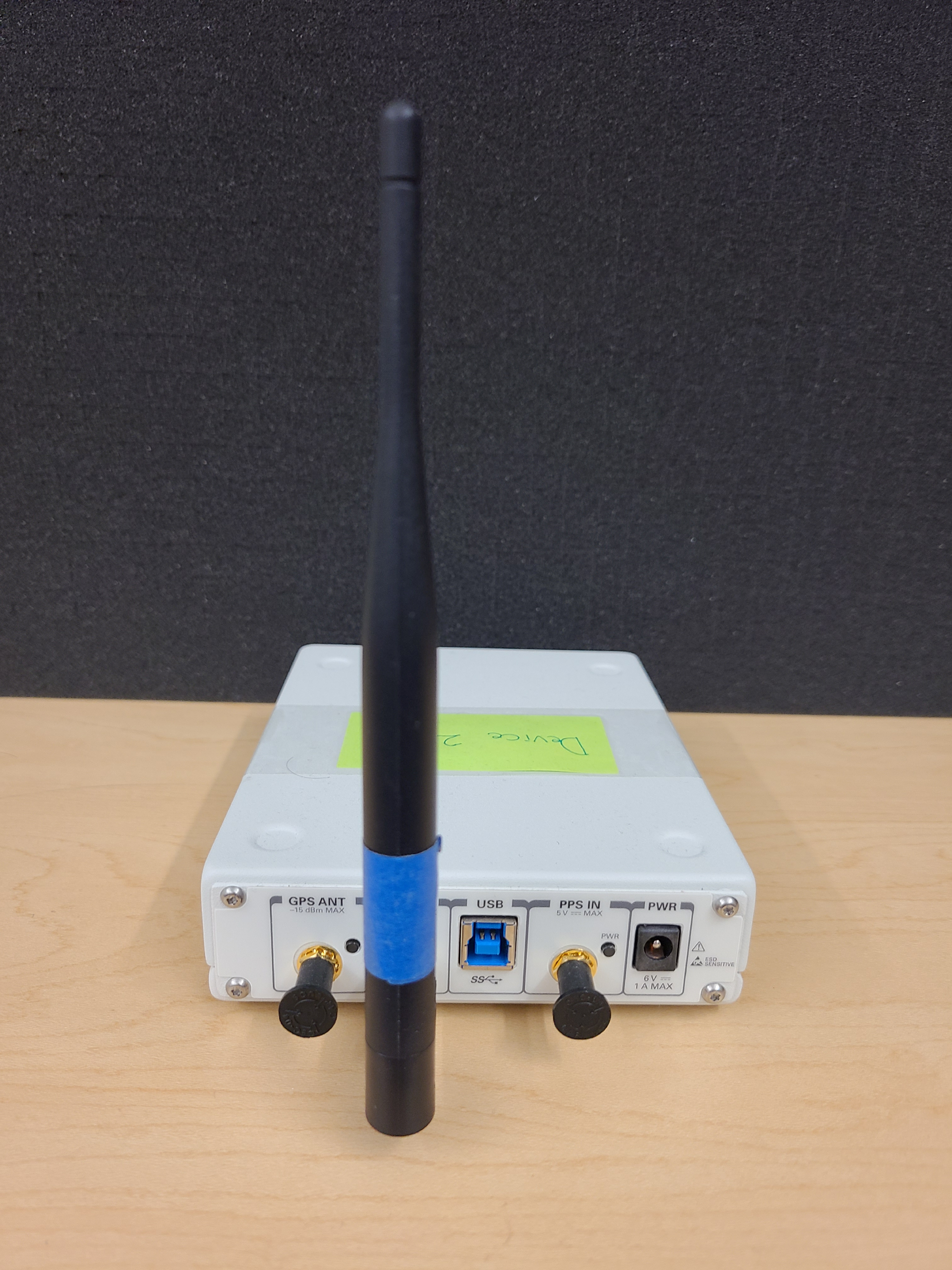}
    \label{fig:ursp}}
\subfloat[1 Lopy and 1 Fipy transmitters]{
    \includegraphics[width=.655\columnwidth]{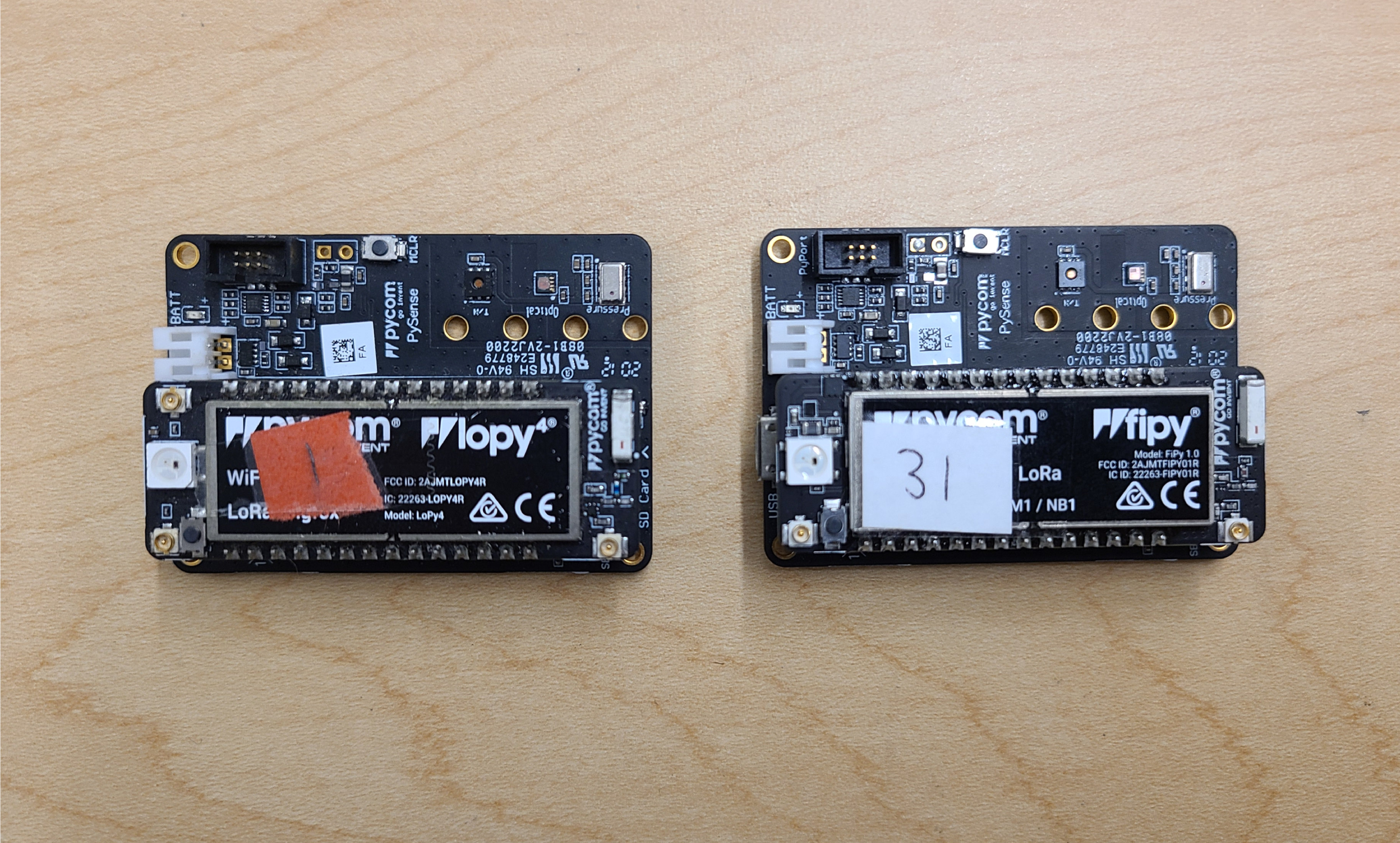}
    \label{fig:pycom}}
\caption{Testbed hardware}
\label{fig:testbed}
\end{figure}

To assess VFDT's robustness, we collected multiple datasets, each from 30 different Pycom devices (13 Lopy and 17 Fipy boards) tested across 5 different locations. Each Pycom device is equipped with a programmable ESP32 that can transmit using WiFi, LoRa and other protocols. An Ettus USRP (Universal Software Radio Peripheral) B210 receiver is used to collect and sample the RF data in the form of raw IQ values through GNURadio. Fig.~\ref{fig:testbed} shows the USRP receiver, as well as a sample of each of the Lopy and Fipy device boards.

The first 3 locations tested are at set distances away from the receiver, all within line of sight of the antenna. Location 1 (Loc 1) is set 1 meter away, while location 2 (Loc 2) is set 2 meters away, and location 3 (Loc 3) is set 3 meters away. Two other considered locations are selected at random from a set of 10 predetermined locations across the testbed area (i.e., an in-door lab environment). These locations vary in distance and angle away from the receiver to vary the channel conditions as much as possible. Later in the evaluation section, Sec.~\ref{sec:results}, these 2 random locations are referred to as "Rand 1" and "Rand 2". 


Each device is plugged into a lipo battery and then given an initial 20 minute warm up period before data collection begins. After warming up, each device is recorded for 2 minutes continuously beginning with Loc 1, followed by Loc 2 and Loc 3, then by Rand 1 and Rand 2. Each device's ESP32 microcontroller is programmed with the same code that transmits the same message repeatedly, with a set delay in between each transmission. Devices are set to use the WiFi protocol at 2.412GHz with a bandwidth of 20MHz. The USRP receiver is set to sample at 45MSps with a gain of 20dBm.

Before any analysis begins, the raw RF data stream captured from each device is processed to remove unnecessary samples and format the actual transmissions properly. The actual transmitted WiFi frames are extracted from the data stream while removing the samples taken during the programmed delays. Then, the WiFi frames are split into their corresponding in-phase (I) and quadrature (Q) signals and then organized into an array. This resulting array of frames is used for multifractal analysis and the accompanying baseline classification.

\section{Device Classification Results}
\label{sec:results}
The testing accuracy is used as the metric for evaluating the learning models, which is the percentage of the correctly classified tested inputs out of the total number of tested inputs. All models were trained for 30 epochs with 90\% of the collected data used for training and the remaining 10\% used for testing. We evaluate both domain/location adaptation and scalability to measure the robustness of the proposed VFDT representation.

\begin{figure}
\centering
\includegraphics[width=\columnwidth]{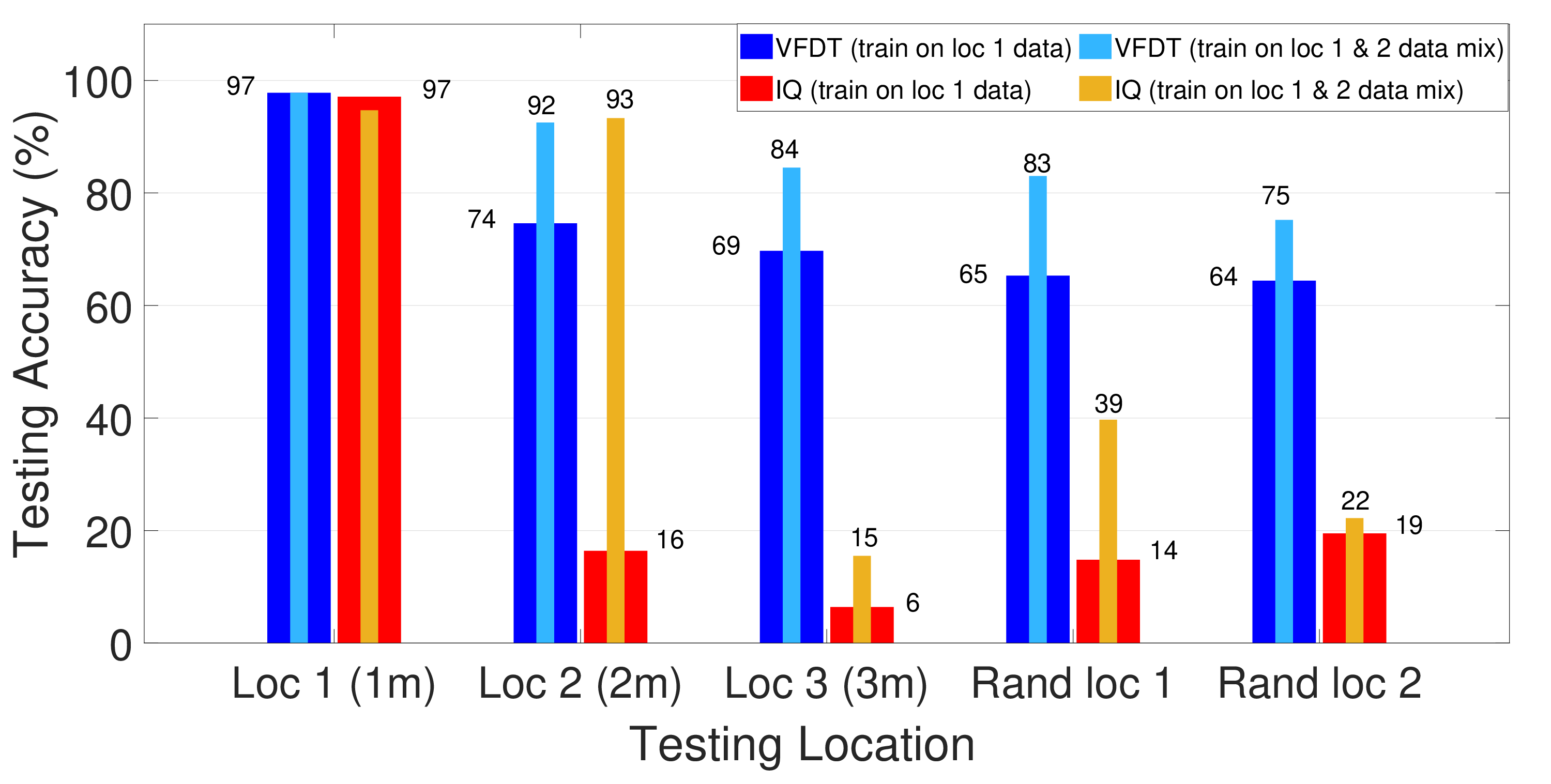}
\caption{Accuracy results when learning models are trained (i) on Loc 1 only data and (ii) on mixed data from both Loc 1 and Loc 2. 
Testing in both scenarios is all done on data from: Loc 1 (1m away), Loc 2 (2m away), Loc 3 (3m away), random location 1 (Rand 1), or Random Location 2 (Rand 2).}
\label{fig:acc}
\end{figure}

\subsection{VFDT Adaptation to Location Changes}

For this evaluation, we considered two scenarios. The first trains the VFDT-based and IQ-based (baseline) deep learning models on data collected at Loc 1 (1m away from the receiver), and test it on data collected from the same location (Loc 1) or from one of the four other locations: Loc 2 (2m away), Loc 3 (3m away), and the two random locations (Rand 1 and Rand 2).
The second scenario consists of training the two learning models on a mixture of data collected from both Loc 1 and Loc 2, and testing it on data collected from one of the 5 considered locations.
The testing accuracy results for these 2 scenarios are shown in Fig. \ref{fig:acc}.
For the first scenario, it can be seen that while the testing accuracy on data captured at Location 1 is similar for both IQ and VFDT approaches, the VFDT model is able to classify data captured from all the other locations at a much higher accuracy level. 
While the IQ model yields an accuracy lower than 20\% for the other 4 locations, the VFDT model yields an accuracy in the high 60\% to low 70\% for the 4 different locations, which is a much smaller drop in performance when changing domains. 

Under the second scenario where the models are trained on data collected from both Locations 1 and 2, 
it can be seen that again the VFDT model outperforms the IQ model. The VFDT model achieves greater results on each of the 3 testing locations (Location 3, Rand 1 and Rand 2) that it was not trained on, once again outperforming the IQ model significantly. Specifically, when tested on Loc 3, Rand 1 and Rand 2, the VFDT model achieves an accuracy of 84\%, 83\% and 74\%, whereas the IQ model achieves 69\%, 65\% and 64\%, respectively.
Comparing these obtained results, it is worth noting that the IQ model performs slightly better on data taken at Location 3, Rand 1 and Rand 2. For example, when tested on data from Location 3, the accuracy of IQ model jumps from 6\% to 15\%. This could be attributed to the benefit of training on combined data collected from two locations, i.e., Locations 1 and 2.
 
In conclusion, our findings have demonstrated that the proposed VFDT representation, when used as an input to deep learning based RF fingerprinting models, yield device features that are more adaptable and generalizable to changing locations.

\subsection{VFDT Scalability with the Number of Devices}
\begin{figure}
\centering
\includegraphics[width=\columnwidth]{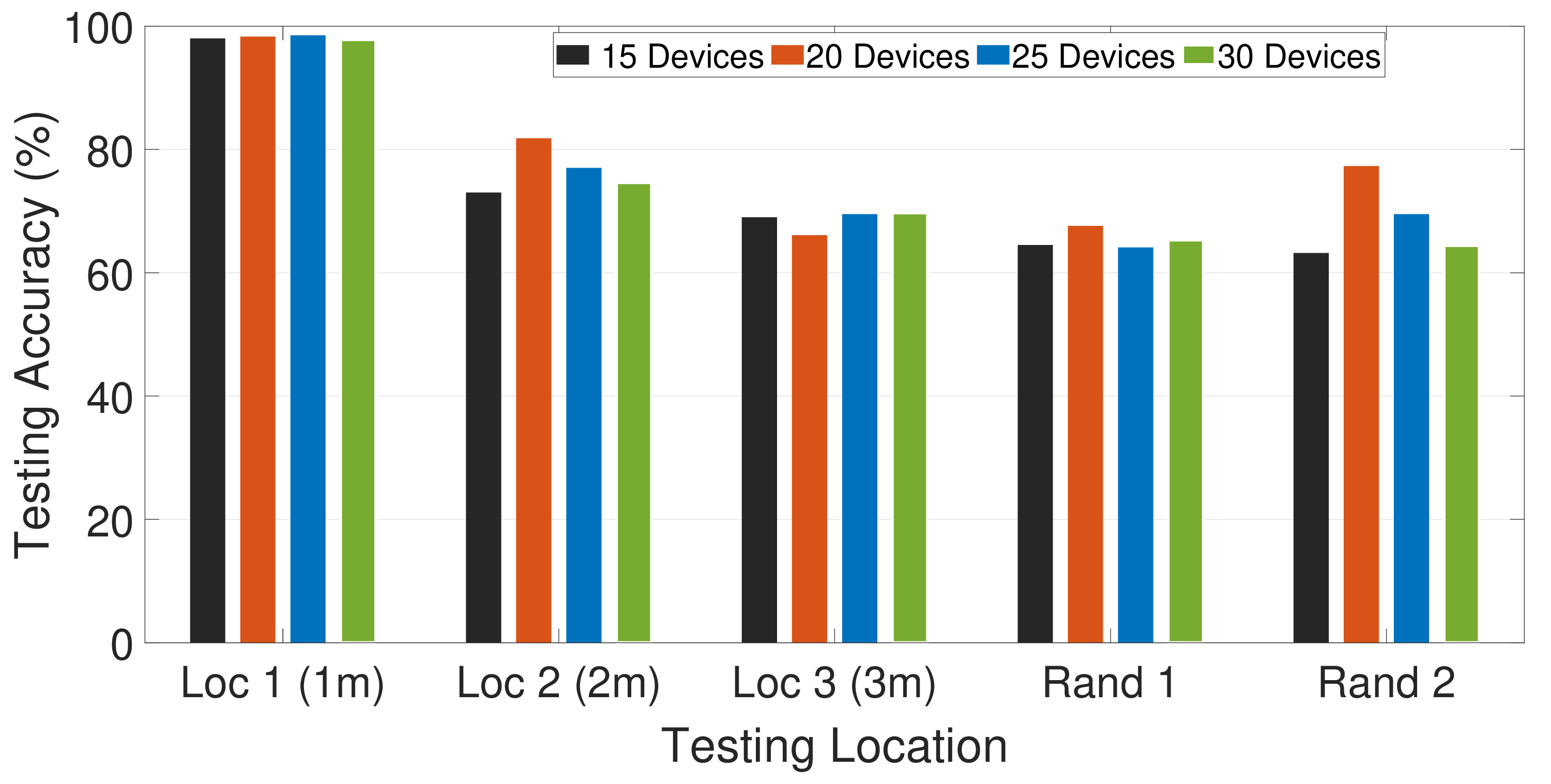}
\caption{Scalability results when models are tested on varying number of devices. All models trained on Location 1 data.}
\label{fig:scale}
\end{figure}

We also looked at how well VFDT scales with the number of devices. 
For this, we trained the models on 5 different subsets of devices, with sizes 15, 20, 25, and 30 devices. For each subset, all models are trained on Loc 1 data (again located 1m away from the receiver) and then tested on all other locations. The resulting testing accuracy for all cases is shown in Fig. \ref{fig:scale}. From the figure, it can be seen that there is no significant drop or falloff in accuracy for a given location across all numbers of devices. Whether the VFDT model is classifying 15 or 30 devices, it is able to do so with very similar performance. Overall, this indicates that the VFDT representation scales well with the number of devices even under varying locations.

\section{Conclusion}
\label{sec:conc}
We use multifractal analysis through the variance fractal dimension trajectory (VFDT) to extract device signatures from RF signals using deep learning. Simulation is performed to analyze how different hardware impairments affect VFDT's ability to separate between devices. It is then demonstrated through experimental datasets that the VFDT representation of the IQ signals is generalizable and more robust than using raw IQ data across varying domains and scales well at classifying wireless devices based on their RF transmissions.

\bibliographystyle{IEEEtran}
\bibliography{References}

\end{document}